\documentclass[conference]{IEEEtran}
\IEEEoverridecommandlockouts
\usepackage{cite}
\usepackage{amsmath,amssymb,amsfonts}
\usepackage{algorithmic}
\usepackage{graphicx}
\usepackage{textcomp}
\usepackage{xcolor}

\usepackage{multirow}
\usepackage{array} 
\usepackage{makecell}
\usepackage[colorlinks, linkcolor=red, anchorcolor=red, citecolor=green]{hyperref}

\def\BibTeX{{\rm B\kern-.05em{\sc i\kern-.025em b}\kern-.08em
    T\kern-.1667em\lower.7ex\hbox{E}\kern-.125emX}}
\begin{document}

\title{Identity-Assisted Association of Unordered DOA Estimates for Neural Speech Source Tracking\\
\thanks{*Corresponding author}
\thanks{
This work was supported by the National Natural Science Foundation of China under Grant 62503399.}
}

\author{
\IEEEauthorblockN{
Bing Yang$^{1,2}$ \qquad
Di Liang$^{1,3}$ \qquad
Xiaofei Li$^{1,4,*}$
}
\IEEEauthorblockA{
$^{1}$ 
School of Engineering, Westlake University, Hangzhou, China\\
$^{2}$ 
School of Artificial Intelligence, Tianjin University, Tianjin, China\\
$^{3}$ 
Zhejiang University, Hangzhou, China\\
$^{4}$ 
Westlake Institute for Advanced Study, Hangzhou, China \\
bingyang@tju.edu.cn, liangdi@westlake.edu.cn, lixiaofei@westlake.edu.cn
}
}

\maketitle

\begin{abstract}
Tracking speech sources remains a challenge due to ambiguous data association arising from intermittent speech, close spatial proximity, and complex acoustic conditions. To address these issues, we propose an identity-assisted association that maps unordered direction-of-arrival (DOA) estimates to speaker-consistent source trajectories for reliable speech source tracking. Specifically, speaker identity embeddings are directly integrated into the model input as a complementary cue to spatial features. This enables maintaining identity consistency by combining long-term time-invariant vocal identity characteristics with the short-term continuity of spatial cues. To effectively process these heterogeneous inputs while accommodating their distinct characteristics, we design a unified neural tracker. Within this model, time self-attention modules capture the temporal evolution of each source, while source self-attention modules distinguish between competing source tracks. Experimental results demonstrate the superiority of the proposed neural tracker in mitigating association confusion for speech source tracking. 

\end{abstract}

\begin{IEEEkeywords}
Speech source tracking, data association, unordered DOA estimates, speaker embedding, self-attention.
\end{IEEEkeywords}

\section{Introduction}

Speech source tracking aims to determine the spatial motion trajectories of active speakers. Capturing the spatial dynamics of speech sources is crucial for real-world applications such as robot audition \cite{DPRTF21}, and audio signal processing tasks like source separation \cite{SESSOverview18}.  Unlike speech source localization, which only predicts the positions and activities of sources without considering their identity association, tracking requires determining not only the location and activity but also the identity of each source \cite{DiaTra23}. Therefore, associating locations and activities with the same speaker is a key problem for speech source tracking. 

Sound source tracking systems typically leverage two complementary cues to ensure reliable association of spatial observations: spatial location cues and sound identity cues. 
Spatial cues offer short-term continuity, reflecting the time-varying trajectories of speakers. They are highly effective when speakers are continuously active and spatially distinct. Sound identity cues capture the time-invariant sounding characteristics, offering long-term consistency that becomes crucial when spatial cues fail. Such failures often occur during intermittent sounding where spatial continuity is broken by silence, when sources are in close proximity, or when spatial features are corrupted by noise and reverberation. 

Based on the cues utilized, existing deep learning-based tracking methods can be categorized into two groups: spatial-only methods \cite{RNNsPIT23,PIRNNsPIT23,Cross3D21,NeuSRP24} which rely exclusively on spatial location cues, and identity-assisted methods \cite{SELDOverview21,SELD19,ACCDOA21,MultiACCDOA22,DiffTrack21,CSTFormer24,Conformer25,C-Conformer26} which explicitly or implicitly incorporate speaker identity information alongside spatial information. Sound source tracking approaches employ various architectures, including Transformer \cite{CSTFormer24,Conformer25,C-Conformer26}, convolutional recurrent neural networks (CRNN) \cite{SELD19,DiffTrack21,NeuSRP24}, convolutional neural network (CNN) \cite{Cross3D21}, recurrent neural network (RNN) \cite{RNNsPIT23,PIRNNsPIT23}. In sound event tracking, a fixed model output node is associated with each specific event class, and the same node is used to produce the location-activity representation of the same sound event \cite{SELDOverview21,SELD19,ACCDOA21}. However, this way is ill-suited for tracking multiple sources belonging to the same class, such as multiple speakers. Since individual speakers are typically unseen during training, it is impractical to allocate a distinct output node for every potential speaker. Hence, speaker tracking models must dynamically assign output nodes to different speakers rather than fixing a one-to-one correspondence, enabling the discrimination of both seen and unseen individuals. To achieve this, strategies such as permutation-invariant training (PIT) \cite{MultiACCDOA22,RNNsPIT23, PIRNNsPIT23} or losses based on multi-object tracking metrics \cite{DiffTrack21} are employed. 
Despite these advancements, how to fully leverage deep learning techniques to exploit the complementarity between spatial location and speaker identity information to robustly track speech sources remains an open challenge. 
%Recently, speaker embedding has been used to perform identity reassignment post-tracking after a front-end speaker tracking system \cite{TrackSpk25, TrackSpk25_2}. 
%Though it explicitly explores the long-term temporal consistency of speaker identity information to help speaker tracking, it is not an neural tracker. 

In this work, we propose a neural speech source tracker that transforms unordered direction-of-arrival (DOA) estimates into ordered source trajectories with the assistance of speaker identity embedding. 
While unordered DOA estimates provide short-term continuity essential for associating them with corresponding speakers, they become unreliable when spatial cues are discontinuous or ambiguous. To address this, our method complements spatial cues with time-invariant vocal identity information, by directly integrating speaker embeddings with unordered DOA estimates at the model input.  
In addition, a model architecture is designed to jointly process unordered DOA estimates and speaker embeddings. Motivated by the basic principles of Bayesian tracking \cite{AVTrackOverview}, which emphasize the critical role of interactions between measurements and tracks as well as between current and historical states, the model processes data along both temporal and source dimensions. Specifically, each source track is represented with a self-attention-based time sequence, where the time self-attention module captures the temporal consistency of each source to maintain long-term identity consistency, and a source-sequence self-attention module is employed to distinguish the multiple source tracks. Experiments conducted on both synthetic and real-world recorded datasets demonstrate that the proposed tracker consistently outperforms compared methods.

\begin{figure*}[t]
  \centering
  \includegraphics[width=0.9\linewidth]{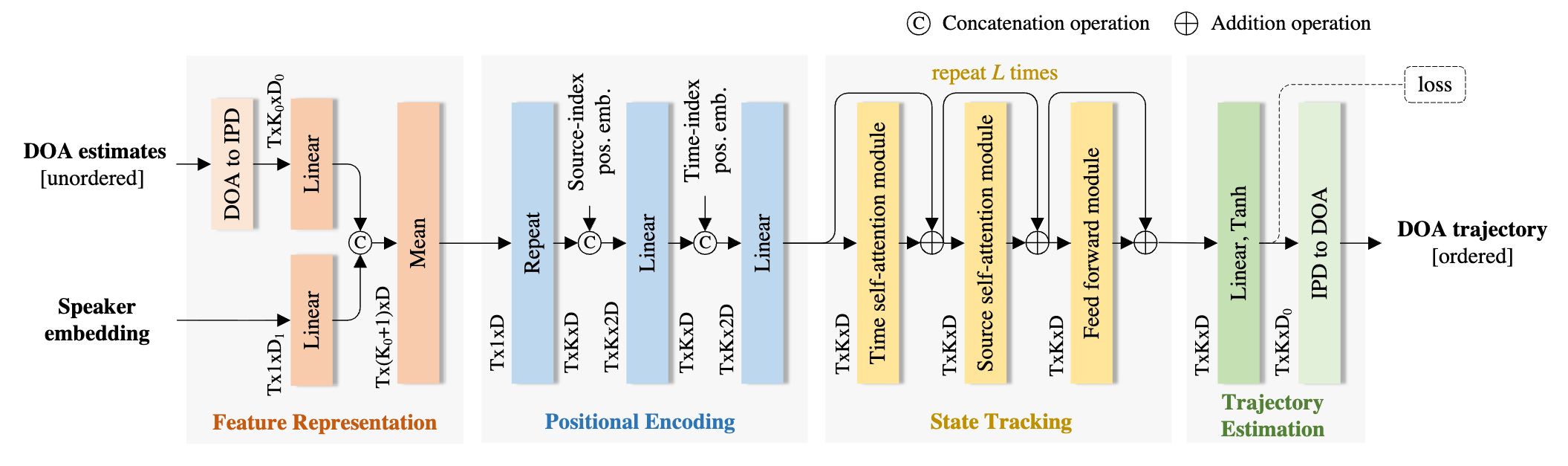}
  \caption{Block diagram of the proposed neural speech source tracker.}
  \label{fig:meth_net}
\end{figure*}

\section{Method}
\label{sec:method}
This section presents the proposed neural speech source tracker, which assists the association of unordered DOA estimates with speaker identity information. As shown in Fig.~\ref{fig:meth_net}, it takes the unordered DOA estimates from a localizer and the speaker embedding from a speaker encoder as input, and outputs the DOA representations/trajectories ordered along the speaker dimension. 

\subsection{DOA representation} 
\label{sec-actipd}
Since speech and many other sound sources are often intermittently silent, the audio signal that contains available DOA/trajectory information will be intermittent. 
Thence, apart from the location representation, an inherent activity indicator is also required. To jointly represent the DOA and the activity information, we use the activity-coupled inter-channel phase difference (IPD) \cite{SRP-DNN22,FN-SSL23}, which encodes the DOA and the activity of a source as: 
\begin{equation}
\label{eq_actipd}
    \mathbf{r}(\boldsymbol{\theta}) = \alpha \times [\mathbf{r}_{21}^T(\boldsymbol{\theta}), \ldots, \mathbf{r}_{M1}^T(\boldsymbol{\theta})]^T \in \mathbb{R}^{2F(M-1)\times1},
\end{equation}
where $\alpha$ and $\boldsymbol{\theta}$ are the activity and the DOA of the potential sound source, which are both time-varying and source-dependent.  $[\mathbf{r}_{21}^T(\boldsymbol{\theta}), \ldots, \mathbf{r}_{M1}^T(\boldsymbol{\theta})]^T$ is the theoretical IPD vector of DOA $\boldsymbol{\theta}$ for a microphone array with microphones $m\in[1, M]$, and 
\begin{equation}
\begin{aligned}
   \mathbf{r}_{mm'}(\boldsymbol{\theta}) &=  [\cos\left(\omega_1\tau_{mm'}(\boldsymbol{\theta})\right),\sin\left(\omega_1\tau_{mm'}(\boldsymbol{\theta})\right),\ldots,\\
   &\cos\left(\omega_F\tau_{mm'}(\boldsymbol{\theta})\right),\sin\left(\omega_F\tau_{mm'}(\boldsymbol{\theta})\right)]^T \in \mathbb{R}^{2F\times1}, 
    \label{eq_dpipd}
\end{aligned}
\end{equation}
where $F$ represents the number of frequencies, and $\omega_f$ denotes the angular frequency of the $f$-th frequency with $f$$\in$$[1, F]$. $\tau_{mm'}(\boldsymbol{\theta})$ is the time difference of arrival (TDOA) between signals captured by the $m$-th and the $m'$-th microphones, which can be computed according to the far-field model \cite{SRP-DNN22}. 

To transform the predicted activity-coupled IPD representation $\tilde{\mathbf{s}}(\boldsymbol{\theta})$ back to DOA and activity, we compute the similarity between this representation with the theoretical IPD vectors of candidate DOAs. Note that the candidate DOAs are uniformly sampled from the whole localization/tracking space. The output DOA is the candidate that has the maximum similarity, and the output activity is the corresponding similarity value. 

\subsection{Network architecture}
The proposed neural tracker is composed of multiple DOA tracks, with each track representing an ordered DOA trajectory (with activity information) for one potential source. The number of tracks $K$ defines the maximum number of sources that can be processed over the whole time.
To update the DOA track of the $k$-th source at the $t$-th time frame, the following information is crucial for maintaining the identity consistency of the same track. 
1) The historical information of $k$-th source facilitates the information interaction over time. 
Specifically, the historical DOA trajectory can help catch the temporal evolution of speaker motion and smooth the DOA trajectory of this source. In addition, the historical information of the speaker embedding can help build the long-time correspondence of one speaker, especially when the speaker is silent for a period before restarting speaking. 
2) Information of other tracks at the $t$-th time frame enhances the information interaction across sources, and helps the one-to-one association of unordered DOA estimates (and speaker embedding) to DOA tracks. 
Hence, the proposed tracker is designed to mainly process data in two sequence dimensions, namely the time dimension and the source/track dimension.
The proposed neural tracker consists of four blocks, namely feature representation, positional encoding, state tracking, and trajectory estimation. 

\textit{Feature representation:} 
The unordered DOA estimates and their activities are first transformed to the activity-coupled IPD representation according to Eq.~(\ref{eq_actipd}). 
The dimension of the IPD representation is $T \times K_{0} \times D_{0}$, where $T$ denotes the number of time frames, $K_{0}$ is the maximum number of active sound sources (and unordered DOA estimates) at each time, and $D_{0}$ is the feature dimension. We set $K_{0}=2$ in this work, which means at most two active speakers are allowed at one time. 
The speaker embedding has a dimension of $T \times 1\times D_{1}$, where $D_{1}$ is its feature dimension. The $K_{0}$ IPD features and the speaker embedding are projected to the same feature dimension $D$ with two separate linear layers, and then are all averaged to get a fused feature with a dimension of $T \times 1 \times D$. The linear projection followed by the average operation serves as a simple yet effective way to jointly represent DOA, activity, and speaker identity at the frame level. 

\textit{Positional encoding:}   
The $K_{0}$ IPD features and the speaker embedding should be properly associated with $K$ DOA tracks ($K_{0} \leq K$). 
The same feature is input to all tracks, and the track networks extract association information from the same feature. To discriminate between different tracks, the $K$-repeated fused features are concatenated with a track-index positional embedding along the feature dimension, which is followed by a linear layer to project to the dimension $D$. To discriminate among different time frames, we also concatenate the time-index positional embedding along the feature dimension, and use a linear layer to compress to the dimension $D$. 

\textit{State tracking:} 
The state tracking block is a cascade of a time self-attention module for temporal information processing of each source track, a source self-attention module for source information interaction at each frame, and a feed-forward module.
The time self-attention and source self-attention modules follow the standard multi-head self-attention mechanism \cite{Transformer17}, which also involves a residual connection and layer normalization. The time self-attention module uses a binary mask to ensure causality, which facilitates the online implementation of sound source tracking. The state tracking block can jointly process the heterogeneous spatial and identity information, fusing these cues to enable reliable speech source tracking. 

\textit{Trajectory estimation:}
A linear layer with an activation of the tanh function is used to transform the output of the state tracking module into an activity-coupled IPD representation of $T \times K \times D_{0}$, which can be transformed back to DOAs and activities following the principle presented in Section \ref{sec-actipd}.  

\subsection{Model input}
Under ideal conditions with continuous speech, spatial location information exhibits short-term continuity, which is crucial for associating unordered DOA estimates with their corresponding sources. However, in real-world scenarios, audio-based location cues often become ambiguous, and tracking based solely on unordered DOA estimates could be infeasible. To reduce the association ambiguity, we incorporate unified-format speaker embedding at the model input as a complementary cue to unordered DOA estimates. 

\textit{Unordered DOA estimates:} The microphone signals are fed into a pretrained localization model, namely IPDnet \cite{IPDnet24}, to predict the time-wise unordered DOAs and activities of sound sources. 
The IPDnet consists of two full-narrow network blocks. The hidden size is 256. To guarantee low complexity, frequency and time dimensions are compressed by factors of 8 and 5 via average pooling layers following the first full-band and second narrow-band modules, respectively. This reduces the input frame rate from 50 frames per second (fps) to 10 fps. The IPDnet is pretrained with 4-second signals using frame-level permutation-invariant training (fPIT) loss. IPDnet outputs $K_0$=2 unordered DOA estimates at each frame, with each DOA estimate corresponding to an active speaker or a null speaker (encoded with a special activity indicator). 

\textit{Speaker embedding:}
The single-channel microphone signal is passed into a pretrained speaker diarization model, namely the causal encoder of the LS-EEND model \cite{EEND25}, to predict time-wise speaker embedding vectors. The encoder of the LS-EEND model mainly consists of four embedding encoder blocks. The hidden size is 256. The output of the embedding encoder has a 10 fps rate. The original LS-EEND model is fine-tuned with 30-second signals used in this work, using the appearance-order loss presented in \cite{EEND25}. We then use the encoder of the fine-tuned LS-EEND for speaker embedding extraction. 
This extracted embedding encodes time-invariant identity information for active speakers across the entire sequence, explicitly indicating source activity and defining the set of candidate speakers capable of generating the observed DOA estimates. It is particularly beneficial when the number of simultaneously active speakers is fewer than the maximum system capacity, as it effectively narrows the search space for data association. 

\subsection{Model training}
The proposed tracker is trained with the mean squared error of the activity-coupled IPD representation between the tracker output and the ground truth. Considering the source/track permutation problem, the permutation-invariant training (PIT) loss is used. To guarantee global identity consistency, the PIT loss is implemented at the utterance level (uPIT).   

\section{Experiments and Discussions}
\subsection{Dataset}
\subsubsection{Synthetic dataset} 
A number of rectangular rooms are simulated using an implementation of the image method \cite{Image_method79} provided by the gpuRIR toolbox\footnote{\url{https://github.com/DavidDiazGuerra/gpuRIR}} \cite{gpuRIR20}. The room size is in the range from 3$\times$3$\times$2.5 m$^3$ to 10$\times$8$\times$6 m$^3$. The reverberation time ranges from 0.2 s to 1.3 s. A 4-channel planar near-rectangular microphone array is placed inside the room, which has the same topology as the microphones 5, 8, 11 and 12 of the robot head array in the localization and tracking (LOCATA) challenge dataset \cite{LOCATA18}. The omnidirectional sound source is located in the same horizontal plane as the microphone array, which can be static in the same place or moving along a series of random sinusoidal continuous trajectories \cite{Cross3D21}.
Speech recordings are selected from the LibriSpeech corpus \cite{LibriSpeech15}. We set at most 2 speakers for each utterance. Multiple speakers are sounding independently. 
The overall speech overlap ratio is about 30 \%.
An arbitrary noise field generator\footnote{\url{https://github.com/ehabets/ANF-Generator}} is used to generate the spatially diffuse white and babble noise \cite{Diffuse08}. 
The microphone signal is created by filtering speech recordings with RIRs, then summing filtered signals of multiple speech sources, and finally scaling and adding diffuse noise with a signal-to-noise ratio (SNR) ranging from -5 dB to 25 dB. 
Each signal is a random combination of the aforementioned data settings regarding source trajectory, microphone position, source signal, noise signal, SNR, room size, reverberation time, absorption coefficients, etc. 
We randomly generate 102,400, 1,024 and 1,024 30-second signals for training, validation and test, respectively. 

\subsubsection{Real-world dataset}
The LOCATA corpus \cite{LOCATA18} is a comprehensive benchmark for sound source localization and tracking, featuring multi-microphone recordings of both static and moving speakers in real-world acoustic environments. 

% \begin{table}[t]
%  \caption{Performance comparison on synthetic data}
%   \label{tab:simu_4spk}
%   \centering
%   \renewcommand\arraystretch{1.2}
%   \tabcolsep0.02in
%   \begin{tabular}{lcccccccccccccc}
%     \hline
%     \multirow{2}{*}{Method} &MDR$\downarrow$ &FAR$\downarrow$ &MAE$\downarrow$  &IDSR$\downarrow$ &AssA$\uparrow$ &DetA$\uparrow$ &HOTA$\uparrow$ \\
%     &[$\%$]  &[$\%$]  &[$^{\circ}$]   &[$\%$]  &[$\%$] &[$\%$]  &[$\%$] \\
%     \hline
%     CST-Former \cite{CSTFormer24}  
%     &21.0 &16.0 &3.8 &1.0 &61.7 &69.3 &64.8 \\
%     C-Conformer \cite{C-Conformer26}  
%     &11.0 &10.1 &3.3 &1.0 &75.7 &81.9 &78.3 \\
%     RNN \cite{RNNsPIT23}
%     &12.7 &8.3 &2.9 &1.0 &60.5 &81.0 &69.1 \\
%     Neural-SRP \cite{NeuSRP24}
%     &17.9 &14.6 &3.3 &1.2 &63.1 &72.9 &67.2  \\
%     \textbf{TR (proposed)}
%     &7.9 &5.9 &2.1 &1.4 &79.9 &87.2 &83.1 \\
%     \hline
%    \end{tabular}
% \end{table}
\begin{table}[t]
 \caption{Performance comparison on synthetic data}
  \label{tab:simu_2spk}
  \centering
  \renewcommand\arraystretch{1.2}
  \tabcolsep0.05in
  \begin{tabular}{lcccccccccccccc}
    \hline
    \multirow{2}{*}{Method} &MDR$\downarrow$ &FAR$\downarrow$ &MAE$\downarrow$  &AssA$\uparrow$ &DetA$\uparrow$ &HOTA$\uparrow$ \\
    &[$\%$]  &[$\%$]  &[$^{\circ}$]   &[$\%$]  &[$\%$] &[$\%$] \\
    \hline
    CST-Former \cite{CSTFormer24}  
    &21.0 &16.0 &3.8 &61.7 &69.3 &64.8 \\
    C-Conformer \cite{C-Conformer26}  
    &11.0 &10.1 &3.3 &75.7 &81.9 &78.3 \\
    RNN \cite{RNNsPIT23}
    &12.7 &8.3 &2.9 &60.5 &81.0 &69.1 \\
    Neural-SRP \cite{NeuSRP24}
    &17.9 &14.6 &3.3 &63.1 &72.9 &67.2  \\
    \textbf{TR (proposed)}
    &\textbf{7.9} &\textbf{5.9} &\textbf{2.1} &\textbf{79.9} &\textbf{87.2} &\textbf{83.1} \\
    \hline
   \end{tabular}
\end{table}

\subsection{Experimental Setup}
\subsubsection{Model configurations and training details}
The sampling rate of the microphone signals is 16 kHz. The window length is 32 ms with a frame shift of 10 ms. 
The number of frequencies $F$ is 256. The number of microphones $M$ is 4.
The maximum number of active sources $K$ for each utterance is set to 2.
The tracker input and output have a 10 fps frame rate. 
The number of tracking blocks $L$ is set to 4. 
The hidden sizes $D$, $D_0$ and $D_1$ are 256, 1536 and 256, respectively. 
The model is trained using the Adam optimizer with mini-batches of 16. The learning rate is initially set to 0.0001 with a cosine-decay scheduler. The maximum number of training epochs is 30. The best model is the one with the minimum validation loss. 

\subsubsection{Evaluation metrics}
The accuracy of detection, association and localization is crucial for sound source tracking.
Higher order tracking accuracy (HOTA) \cite{HOTA21, TrackSpkData25} is a single unified metric that equally weights detection accuracy (DetA) and association accuracy (AssA). We use HOTA as the principal indicator for overall tracker accuracy. Other metrics including DetA, AssA, miss detection rate (MDR), false detection rate (FAR), and mean absolute error (MAE) are provided for deeper insights into tracker behaviors. 
When computing the detection-related metrics, the localization (DOA error) threshold is set to 10$^{\circ}$. 
Miss detection (MD) refers to source active but not detected, and false alarm (FA) means source detected but not active. The MDR and the FAR are computed as the percentage of MDs and FAs out of the active sources of all time frames, respectively \cite{LOCATA_Results20,SRP-DNN22}. 
The MAE is computed by averaging the absolute azimuth error of all successfully localized sources and time frames. 

\subsubsection{Comparison methods}
The proposed tracker is compared with four tracking methods: CST-Former \cite{CSTFormer24}, C-Conformer \cite{C-Conformer26}, RNN \cite{RNNsPIT23}, and Neural-SRP \cite{NeuSRP24}. For fair comparison, all models are adapted to use synthetic microphone signals and trained on the same synthetic dataset.
% For CST-Former and C-Conformer, we adapted the SELD architectures originally designed for Ambisonics, to accept real and imaginary spectral components instead of their native log-Mel and intensity features. For the RNN baseline, we feeding it unordered ACC-DOA features derived from the same DOA measurements as our framework. Besides, we modified Neural-SRP, which typically employs a differentiable tracking loss, to minimize the MSE of ACC-DOA representations, thereby aligning its optimization objective with the other baselines.
% namely the recurrent neural network (RNN)-based model in \cite{RNNsPIT23}. It uses RNN to process and output ACCDOA-format \cite{ACCDOA21} spatial representations. 
%For fair comparison, all methods use the same unordered DOA estimates, which are output by the localization model (namely IPDnet \cite{IPDnet24}) and then randomly shuffled along the source dimension.

\subsection{Experimental Results}

\subsubsection{Comparison on synthetic data} 

Table~\ref{tab:simu_2spk} presents the tracking results on synthetic data, where the proposed transformer-based model is denoted as TR. 
Our method shows an advantage over all comparison approaches.  
When compared to identity-assisted methods, namely C-Conformer and CST-Former, our method achieves superior performance primarily due to its more robust tracking framework. 
Our method also outperforms these spatial-only methods, namely RNN and Neural-SRP, by a wide margin. This substantial gain is attributed to both our network architecture and the explicit integration of speaker identity cues. These findings confirm the effectiveness of the proposed tracking method. 

\begin{table}[t]
 \caption{Performance comparison on the LOCATA dataset}
  \label{tab:locata}
  \centering
  \renewcommand\arraystretch{1.2}
  \tabcolsep0.055in
  \begin{tabular}{lcccccccccccccc}
  \hline
    \multirow{2}{*}{Method} &MDR$\downarrow$ &FAR$\downarrow$ &MAE$\downarrow$  &AssA$\uparrow$ &DetA$\uparrow$ &HOTA$\uparrow$ \\
    &[$\%$]  &[$\%$]  &[$^{\circ}$]   &[$\%$] &[$\%$]  &[$\%$] \\
    \hline
    CST-Former \cite{CSTFormer24}  
    &24.2 &24.2 &4.2 &44.8 &62.9 &52.4\\
    C-Conformer \cite{C-Conformer26} 
    &20.1 &18.1 &4.2 &55.5 &69.1 &60.9\\
    RNN \cite{RNNsPIT23} 
    &10.5 &8.4 &3.1 &61.5 &83.0 &69.9\\
    Neural-SRP \cite{NeuSRP24}  
    &28.8 &25.6 &3.9 &45.0 &56.6 &49.4 \\
    \textbf{TR (proposed)} 
    &\textbf{9.3} &\textbf{5.9} &\textbf{2.3} &\textbf{70.0} &\textbf{85.8} &\textbf{76.4}\\
    \hline
   \end{tabular}
\end{table}

\begin{table}[t]
 \caption{Ablation study on proposed tracker}
  \label{tab:simu_ablation}
  \centering
  \renewcommand\arraystretch{1.2}
  \tabcolsep0.025in
  \begin{tabular}{lcccccccccccccc}
    \hline
    \multirow{2}{*}{Method} &MDR$\downarrow$ &FAR$\downarrow$ &MAE$\downarrow$  &AssA$\uparrow$ &DetA$\uparrow$ &HOTA$\uparrow$ \\
    &[$\%$]  &[$\%$]  &[$^{\circ}$]   &[$\%$]  &[$\%$] &[$\%$]  \\
    \hline
    \makecell[l]{Unordered DOA estimates\\ (IPDnet \cite{IPDnet24})}
    % &11.3 &2.7 &2.2 &2.1 &54.8 &86.5 &68.0 \\ % th=0.55
    &9.0 &6.1 &2.2 &54.2 &86.0 &67.4 \\ % th=0.235
    TR w/o spk self-attention
    &8.3 &6.5 &\textbf{2.1} &79.4 &86.4 &82.5\\
    TR w/o time self-attention
    % &11.0 &5.0 &2.1 &3.1 &60.7 &85.0 &70.9\\ th=0.5
    &10.1 &6.6 &\textbf{2.1 }&60.2 &84.7 &70.5\\ %th=0.48
    TR w/o spk emb.
    %&9.8 &4.9 &2.1 &1.3 &73.7 &86.3 &79.0\\ % th=0.5
    &9.5 &6.1 &\textbf{2.1} &73.2 &85.7 &78.5\\ % th=0.44
    \textbf{TR (proposed)}
    &\textbf{7.9} &\textbf{5.9} &\textbf{2.1} &\textbf{79.9} &\textbf{87.2} &\textbf{83.1} \\ % th=0.4
    \hline
   \end{tabular}
\end{table}

\begin{figure*}[t]
  \centering
  \includegraphics[width=0.83\linewidth]{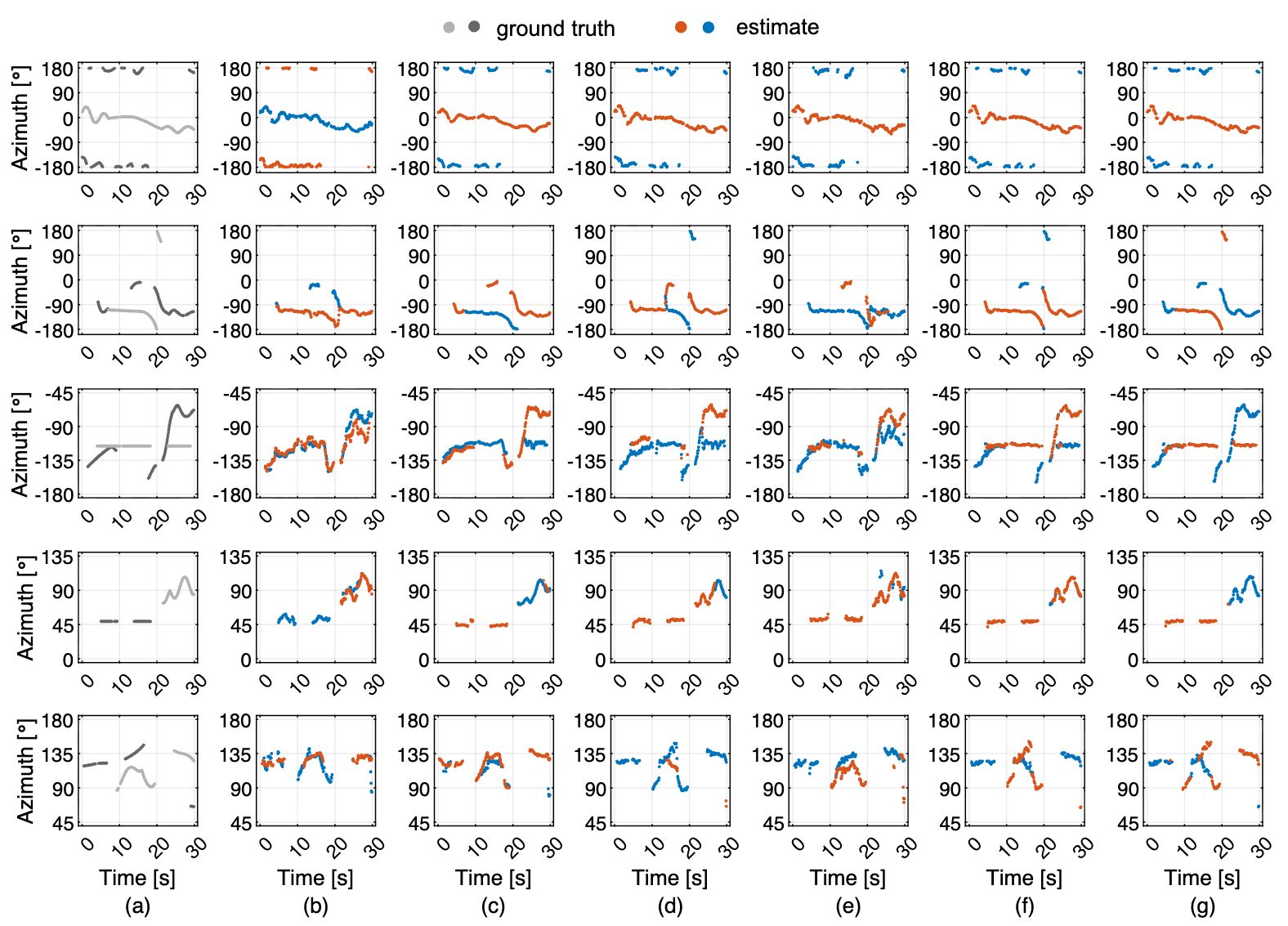}
  \caption{Illustration of DOA trajectory for (a) ground truth, (b) CST-Former \cite{CSTFormer24}, (c) C-Conformer \cite{C-Conformer26}, (d) RNN \cite{RNNsPIT23}, (e) Neural-SRP \cite{NeuSRP24}, (f) proposed tracker w/o speaker embedding, and (g) proposed tracker w/ speaker embedding. Each row corresponds to a recording example. Different colors indicate different speaker identities.} 
  \label{fig:exp}
\end{figure*}

\subsubsection{Comparison on LOCATA dataset} 

Table~\ref{tab:locata} presents the instance-level averaged results on tasks 3-6 of the LOCATA dataset. For evaluation, we use the model trained on 30-second synthetic data and then fine-tuned on 60-second synthetic data. The models trained on the synthetic dataset are directly evaluated on the LOCATA dataset.  
The results demonstrate that our proposed method consistently outperforms all competing approaches, which is consistent with that observed on the simulated dataset. These findings validate the robustness of our approach and confirm its feasibility for generalization to real-world acoustic data. 

\subsubsection{Ablation study on proposed tracker}

To figure out the contribution of each model part, ablation studies are carried out on the proposed tracker under the condition $K$=2. The results are shown in Table~\ref{tab:simu_ablation}. To demonstrate the benefit of the tracker over the localizer, we also present the performance of the unordered DOA estimates obtained from the localization model \cite{IPDnet24} as a reference. 
Although speaker identity-related metrics have not been used to evaluate the localization method, it still provides an indication of the ordering degree (i.e., association accuracy) of the DOA estimates. 
It can be seen that removing either the speaker or time self-attention modules degrades performance, with the latter causing a more significant drop in AssA, highlighting its necessity for long-term identity consistency. 
Moreover, the tracker without speaker embeddings outperforms unordered DOA inputs in terms of association accuracy, despite a slight trade-off in detection precision. Incorporating speaker embeddings further improves both association accuracy and detection accuracy. 
These results demonstrate that the proposed model not only effectively leverages spatial information for robust tracking, but also successfully integrates speaker identity cues to enhance overall performance. The superiority of our tracker over unordered DOA estimates confirms its ability to ensure long-term identity consistency and refine noisy DOA trajectories. 

\subsubsection{Trajectory visualization}

To better understand the behaviors of these trackers, five  tracking examples are visualized in Fig.~\ref{fig:exp}. 
In the first scenario, where two sources are continuously active and spatially well-separated, all methods achieve successful tracking.
In the second scenario, involving one intermittently active source, only C-Conformer and the proposed method with speaker embeddings successfully complete the task. Notably, the proposed method demonstrates superior accuracy compared to C-Conformer.
In the third scenario, characterized by closely located speakers and trajectory crossings, CST-Former and Neural-SRP exhibit low detection and association accuracy, failing to resolve the sources. The proposed method shows better detection accuracy, and using speaker embedding yields the best performance. 
In the fourth scenario, where two speakers alternate speaking turns, tracking fails for all comparison methods due to association ambiguity, whereas the proposed method with speaker embedding remains robust.  
In the fifth scenario, which involves intermittent speech and closely located overlapping sources, the trajectory condition becomes highly complex. Consequently, all methods exhibit degraded performance under these challenging conditions. 
Overall, the comparison between the proposed trackers with and without speaker embeddings reveals that incorporating identity cues effectively mitigates identity confusion, particularly in scenarios involving intermittent speech activity, close spatial proximity, or take-turns talking.
 
\section{Conclusion}
\label{sec:conlu}

This paper presents an identity-assisted neural speech source tracker designed to transform unordered DOA estimates into ordered source trajectories. 
A key feature of our approach is the direct integration of speaker embeddings into the model input without necessitating extensive architectural modifications. This improves temporal identity consistency, particularly in challenging conditions with ambiguous spatial cues. 
By employing an alternating mechanism of time and source self-attention modules, the proposed framework effectively associates unordered DOA estimates with existing tracks. 
Experimental results show that the identity-assisted association improves trajectory consistency, and the proposed tracker achieves better tracking performance than the compared methods in dynamic scenarios. 
This study focuses on two-speaker tracking with at most two active sources per frame. 
In the future, we will scale the system to accommodate a larger number of speakers and process longer temporal sequences in increasingly complex acoustic environments.

\bibliographystyle{IEEEtran}
\bibliography{mybib}

@STRING{IEEE_J_ASLP_new   = "IEEE/ACM Trans. Audio, Speech, Lang. Process."}

@STRING{IEEE_J_ASLPRO     = "IEEE Trans. Audio, Speech Lang. Process."}

@STRING{IEEE_J_STSP       = "IEEE J. Selected Topics Signal Process."}

@STRING{IEEE_OJ_SP        = "IEEE Open J. Signal Process."}

@STRING{J_ASA             = "J. Acoust. Soc. Amer."}

@STRING{IJCV              = "Int. J. Comput. Vis."}

@STRING{IEEE_ICASSP       = "Proc. IEEE Int. Conf. Acoust., Speech, Signal Process."}

@STRING{IEEE_WASPAA       = "Proc. IEEE Workshop Appl. Signal Process. Audio Acoust."}

@STRING{IEEE_SAM          = "Proc. IEEE Sensor Array Multichannel Signal Process. Workshop"}

@STRING{EUSIPCO           = "Proc. Euro. Signal Process. Conf."}

@STRING{INTERSPEECH       = "Proc. INTERSPEECH"}

@STRING{DCASE             = "Proc. Detect. and Classification of Acoust. Scenes Events Workshop"}

@STRING{NeurIPS           = "Proc. Int. Conf. Neural Inf. Process. Syst."}

@STRING{IEEE_SLT          = "Proc. IEEE Spoken Language Technol. Workshop"}

@STRING{EAAFA             = "Proc. Eur. Acoust. Assoc. Forum Acusticum"}

@STRING{IEEE_J_ASLP_new   = "IEEE/ACM Transactions on Audio, Speech, and Language Processing (TASLP)"}

@STRING{IEEE_J_ASLPRO     = "IEEE Transactions on Audio, Speech and Language Processing (TASLPRO)"}

@STRING{IEEE_J_STSP       = "IEEE Journal of Selected Topics in Signal Processing"}

@STRING{IEEE_OJ_SP        = "IEEE Open Journal of Signal Processing"}

@STRING{J_ASA             = "Journal of the Acoustical Society of America"}

@STRING{IJCV              = "International Journal of Computer Vision"}

@STRING{IEEE_ICASSP       = "IEEE International Conference on Acoustics, Speech and Signal Processing (ICASSP)"}

@STRING{IEEE_WASPAA       = "IEEE Workshop on Applications of Signal Processing to Audio and Acoustics (WASPAA)"}

@STRING{IEEE_SAM          = "IEEE Sensor Array and Multichannel Signal Processing Workshop"}

@STRING{EUSIPCO           = "European Signal Processing Conference (EUSIPCO)"}

@STRING{INTERSPEECH       = "Annual Conference of the International Speech Communication Association (INTERSPEECH)"}

@STRING{DCASE             = "Detection and Classification of Acoustic Scenes and Events Workshop (DCASE)"}

@STRING{NeurIPS           = "International Conference on Neural Information Processing Systems (NIPS)"}

@STRING{IEEE_SLT          = "IEEE Spoken Language Technology Workshop (SLT)"}

@STRING{EAAFA             = "European Acoustics Association Forum Acusticum"}

@article{DPRTF21,
    Author = "Bing Yang and Hong Liu and Xiaofei Li",
    Title = "Learning Deep Direct-Path Relative Transfer Function for Binaural Sound Source Localization",
    Journal = IEEE_J_ASLP_new,
    Volume = "29",	
    Number = "",
    Pages = "3491-3503",
    Year = "2021"}

@InProceedings{SRP-DNN22,
    author = "Bing Yang and Hong Liu and Xiaofei Li",
    title = "{SRP-DNN}: Learning Direct-Path Phase Difference for Multiple Moving Sound Source Localization",
    booktitle = IEEE_ICASSP,
    year = "2022",
    pages = "721-725"}

@InProceedings{FN-SSL23,
    author = "Yabo Wang and Bing Yang and Xiaofei Li",
    title = "{FN-SSL}: Full-Band and Narrow-Band Fusion for Sound Source Localization",
    booktitle = INTERSPEECH,
    year = "2023",
    pages = "3779-3783"}

@article{IPDnet24,
    Author = "Yabo Wang and Bing Yang and Xiaofei Li",
    Title = "{IPDnet}: A Universal Direct-Path {IPD} Estimation Network for Sound Source Localization",
    Journal = IEEE_J_ASLP_new,
    Volume = "32",	
    Number = "",
    Pages = "5051-5064",
    Year = "2024"}

@article{gpuRIR20,
    Author = "David Diaz-Guerra and Antonio Miguel and Jose R. Beltran",
    Title = "{gpuRIR}: A python library for room impulse response simulation with {GPU} acceleration",
    Journal = "Multimedia Tools Appl.",
    Volume = "80",	
    Number = "4",
    Pages = "5653-5671",
    Year = "2021"}

@article{Image_method79,
    Author = "Jont B. Allen and Daivid A. Berkley",
    Title = "Image method for efficiently simulating small-room acoustics",
    Journal = J_ASA,
    Volume = "65",	
    Number = "4",
    Pages = "943--950",
    Year = "1979"}

@InProceedings{LibriSpeech15,
    author = "Vassil Panayotov and Guoguo Chen and  Daniel Povey and  Sanjeev Khudanpur",
    title = "{LibriSpeech}: An {ASR} corpus based on public domain audio books",
    booktitle = IEEE_ICASSP,
    year = "2015",
    pages = "5206-5210"}

@article{Diffuse08,
    Author = "Emanuel A. P. Habets and Israel Cohen and Sharon Gannot",
    Title = "Generating nonstationary multisensor signals under a spatial coherence constraint",
    Journal = J_ASA,
    Volume = "124",	
    Number = "5",
    Pages = "2911-2917",
    Year = "2008"}

@InProceedings{LOCATA18,
    author = "Heinrich W. Lollmann and Christine Evers and Alexander Schmidt and Heinrich Mellmann and Hendrik Barfuss and Patrick A. Naylor and Walter Kellermann",
    title = "The {LOCATA} challenge data corpus for acoustic source localization and tracking",
    booktitle = IEEE_SAM,
    year = "2018",
    pages = "410-414"}

@article{LOCATA_Results20,
    Author = {Christine Evers and Heinrich W. Lollmann and Heinrich Mellmann and Alexander Schmidt and Hendrik Barfuss and Patrick A. Naylor and Walter Kellermann},
    Title = {The {LOCATA Challenge}: Acoustic Source Localization and Tracking},
    Journal = IEEE_J_ASLP_new,
    Volume = {8},	
    Number = {},
    Pages = {1620--1643},
    Year = {2020}}

@article{HOTA21,
    Author = "Jonathon Luiten and Aljosa Osep and Patrick Dendorfer and Philip Torr and Andreas Geiger and Laura Leal-Taixe and Bastian Leibe",
    Title = "{HOTA}: A higher order metric for evaluating multi-object tracking",
    Journal = IJCV,
    Volume = "192",	
    Number = "2",
    Pages = "548-578",
    Year = "2021"}

@article{EEND25,
    Author = "Di Liang and Xiaofei Li",
    Title = "{LS-EEND}: Long-Form Streaming End-to-End Neural Diarization With Online Attractor Extraction",
    Journal = IEEE_J_ASLPRO,
    Volume = "33",	
    Number = "",
    Pages = "3568-3581",
    Year = "2025"}

@article{Cross3D21,
    Author = "David Diaz-Guerra and Antonio Miguel and Jose R. Beltran",
    Title = "Robust Sound Source Tracking Using {SRP-PHAT} and 3{D} Convolutional Neural Networks",
    Journal = IEEE_J_ASLP_new,
    Volume = {29},	
    Pages = {300-311},
    Year = {2021}}

@InProceedings{DiaTra23,
    author = "Jeremy H. M. Wong and Yifan Gong",
    title = "Joint Speaker Diarisation and Tracking in Switching State-Space Model",
    booktitle = IEEE_SLT,
    year = "2023",
    pages = "605-612"}

@InProceedings{Transformer17,
    author = "Ashish Vaswani and Noam Shazeer and Niki Parmar and Jakob Uszkoreit and Llion Jones and Aidan N Gomez and Łukasz Kaiser and Illia Polosukhin",
    title = "Attention is All you Need",
    booktitle = NeurIPS,
    year = "2017",
    pages = "5998-6008"}

@article{SESSOverview18,
    Author = {Deliang Wang and Jitong Chen},
    Title = {Supervised Speech Separation Based on Deep Learning: An Overview},
    Journal = IEEE_J_ASLP_new,
    Volume = {26},	
    Number = {10},
    Pages = {1702-1726},
    Year = {2018}}

@article{AVTrackOverview,
    Author = "Jinzheng Zhao and Yong Xu and Xinyuan Qian and Davide Berghi and Peipei Wu and Meng Cui and Jianyuan Sun and Philip Jackson and Wenwu Wang",
    Title = "Audio-Visual Speaker Tracking: Progress, Challenges, and Future Directions",
    Journal = "arXiv preprint arXiv:2310.14778",
    Volume = "",	
    Number = "",
    Pages = "",
    Year = "2023"}

@article{SELD19,
    Author = "Sharath Adavanne and Archontis Politis and Joonas Nikunen and Tuomas Virtanen",
    Title = "Sound Event Localization and Detection of Overlapping Sources Using Convolutional Recurrent Neural Networks",
    Journal = IEEE_J_STSP,
    Volume = {13},	
    Number = {1},
    Pages = {34-48},
    Year = {2019}}

@article{SELDOverview21,
    Author = {Archontis Politis and Annamaria Mesaros and Sharath Adavanne and Tuomas Virtanen},
    Title = {Overview and Evaluation of Sound Event Localization and Detection in {DCASE} 2019},
    Journal = IEEE_J_ASLP_new,
    Volume = {29},	
    Number = {},
    Pages = {684-698},
    Year = {2021}}

@InProceedings{ACCDOA21,
    author = "Kazuki Shimada and Yuichiro Koyama and Naoya Takahashi and Shusuke Takahashi and Yuki Mitsufuji",
    title = "{ACCDOA}: Activity-Coupled Cartesian Direction of Arrival Representation for Sound Event Localization And Detection",
    booktitle = IEEE_ICASSP,
    year = "2021",
    pages = "915-919"}

@InProceedings{MultiACCDOA22,
    author = "Kazuki Shimada and Yuichiro Koyama and Naoya Takahashi and Shusuke Takahashi and Yuki Mitsufuji",
    title = "{Multi-ACCDOA}: Localizing And Detecting Overlapping Sounds From The Same Class With Auxiliary Duplicating Permutation Invariant Training",
    booktitle = IEEE_ICASSP,
    year = "2022",
    pages = "316-320"}

@InProceedings{CSTFormer24,
    author = "Yusun Shul and Jung-Woo Choi",
    title = "{CST-Former}: Transformer with Channel-Spectro-Temporal Attention for Sound Event Localization and Detection",
    booktitle = IEEE_ICASSP,
    year = "2024",
    pages = "8686-8690"}

@InProceedings{Conformer25,
    author = "Yuxuan Dong and Qing Wang and Hengyi Hong and Ya Jiang and Shi Cheng",
    title = "An Experimental Study on Joint Modeling for Sound Event Localization and Detection with Source Distance Estimation",
    booktitle = IEEE_ICASSP,
    year = "2025",
    pages = "1-5"}

@InProceedings{C-Conformer26,
    author = "Changjiang He and Siyao Cheng and Jiahua Bao and Jie Liu",
    title = "{C-Conformer}: Channel-Augmented {Conformer} for Sound Event Localization and Detection",
    booktitle = IEEE_ICASSP,
    year = "2026",
    pages = "20626-20630"}

@InProceedings{DiffTrack21,
    author = "Sharath Adavanne and Archontis Politis and Tuomas Virtanen",
    title = "Differentiable Tracking-Based Training of Deep Learning Sound Source Localizers",
    booktitle = IEEE_WASPAA,
    year = "2021",
    pages = "211-215"}

@InProceedings{RNNsPIT23,
    author = "David Diaz-Guerra and Archontis Politis and Tuomas Virtanen",
    title = "Position Tracking of a Varying Number of Sound Sources with Sliding Permutation Invariant Training",
    booktitle = EUSIPCO,
    year = "2023",
    pages = "251-255"}

@InProceedings{PIRNNsPIT23,
    author = "David Diaz-Guerra and Archontis Politis and Antonio Miguel and Jose R. Beltran and Tuomas Virtanen",
    title = "Permutation Invariant Recurrent Neural Networks for Sound Source Tracking Applications",
    booktitle = EAAFA,
    year = "2023",
    pages = "2137-2142"}

@article{NeuSRP24,
    Author = "Eric Grinstein and Christopher M. Hicks and Toon van Waterschoot and Mike Brookes and Patrick A. Naylor",
    Title = "The {Neural-SRP} Method for Universal Robust Multi-Source Tracking",
    Journal = IEEE_OJ_SP,
    Volume = "5",	
    Number = "",
    Pages = "19-28",
    Year = "2024"}

@InProceedings{TrackSpkData25,
    author = "Taous Iatariene12 and Alexandre Guerin and Romain Serizel",
    title = "Tracking of Intermittent and Moving Speakers: Dataset and Metrics",
    booktitle = EAAFA,
    year = "2025",
    pages = "4291-4298"}

\end{document}